\documentclass[conference]{IEEEtran}
\IEEEoverridecommandlockouts
\usepackage{cite}
\usepackage{amsmath,amssymb,amsfonts}
\usepackage{algorithmic}
\usepackage{graphicx}
\usepackage{textcomp}
\usepackage{xcolor}
\def\BibTeX{{\rm B\kern-.05em{\sc i\kern-.025em b}\kern-.08em
    T\kern-.1667em\lower.7ex\hbox{E}\kern-.125emX}}
\begin{document}

\title{Auto-labelling of Bug Report using Natural Language Processing} 
\author{\IEEEauthorblockN{Avinash Patil}
\IEEEauthorblockA{\textit{Juniper Networks Inc.} \\
Sunnyvale, USA \\
patila@juniper.net}

\and

\IEEEauthorblockN{Aryan Jadon}
\IEEEauthorblockA{
\textit{San José State University}\\
San Jose, USA \\
aryan.jadon@sjsu.edu}

}

\maketitle

\begin{abstract}

The exercise of detecting similar bug reports in bug tracking systems is known as duplicate bug report detection. Having prior knowledge of a bug report's existence reduces efforts put into debugging problems and identifying the root cause. Rule and Query-based solutions recommend a long list of potential similar bug reports with no clear ranking. In addition, triage engineers are less motivated to spend time going through an extensive list. Consequently, this deters the use of duplicate bug report retrieval solutions. In this paper, we have proposed a solution using a combination of NLP techniques. Our approach considers unstructured and structured attributes of a bug report like summary, description and severity, impacted products, platforms, categories, etc. It uses a custom data transformer, a deep neural network, and a non-generalizing machine learning method to retrieve existing identical bug reports. We have performed numerous experiments with significant data sources containing thousands of bug reports and showcased that the proposed solution achieves a high retrieval accuracy of 70\% for recall@5. 
\end{abstract}


\section{Introduction}
A bug report, also known as a problem report, usually contains details that help an engineer identify and reproduce issues in software or hardware. The information can be structured or unstructured and the summary or description of the issue constitute further detailed unstructured information. However, impacted products, platforms, categories, etc., can be structured information. Bug reports are managed using Bug Tracking Systems. These systems encounter large volumes of duplicate reports constantly. A thorough root cause analysis is needed before the problem report can be determined as a redundant report of an already known issue. The larger the time gap between two reports, the harder it is to identify the later filed report as duplicate.
Typically, the person triaging a new Bug Report determines if the similar issue at hand has been reported in the past or not. To search for an existing bug report and associated root cause analysis search query can be used to retrieve past similar incidents. The query mechanism requires knowing the exact pattern, string, or substring to look for. The difficulty with the traditional querying approach is that every human describes the similar problem in different manner. The summary and description of a problem vary when reported by different users. This also applies when the same user is reporting the same issue reencountered. The language and keywords used by the same person would change over time. If the database query matches something, it will return all the bug reports containing those patterns, strings, or substrings \cite{b3}. This is not much help to the user as there is no way to determine how similar these existing reports are to the new one unless done manually by going through each result returned by the query. Moreover, identifying a duplicate bug report before it enters the system results in but not limited to; preventing repetitive root cause analysis, avoiding duplicate debug effort, enabling auto-filing bug reports for automated workflows, keeping track of recurring problems, and reducing the redundant data in the database.

Our proposed solution can flag a problem report as a duplicate before it enters the system, which helps reduce operational expenses significantly. It also allows customers to obtain a quicker resolution if the problem has already been reported internally or by third-party stakeholders. Along with preventing duplicates from entering the system \cite{b4}, the proposed solution can also help engineers debug or fix the issue efficiently by auto linking existing identical bug reports to new bug reports.

\subsection{Terminology}
A Bug Report can also be referred to as Problem Report or Defect Report in this and other related studies. A Duplicate or Child bug report means that it has been identified as a clone or duplicate of a previously submitted bug report. The previously submitted report in the bug tracking system can be referred to as a Master, Parent, or Original. Child and Parent reports are mapped to each other. The rest of the reports that have never been linked as Child or Parent to another bug report are called Unique Reports. 

\subsection{Paper Outline}
The rest of the paper is outlined as follows, Section 2 provides related work, and Section 3 describes the methodology in detail. Section 4 reports the evaluation results. Section 5 discusses the limitations of this approach. Section 6 concludes the paper.  

\section{Relevant Work}

Runeson, Alexanderson and Nyholm conducted one of the earliest experiments applying ML (Machine Learning) techniques to resolve the issue of duplicate bug reports. For preprocessing, they used stemming, thesaurus, and spellchecker and removed stop words. This resulted in a solution that can deal with varying descriptions of the same bug report. After the preprocessing steps, bug reports were transformed into vectors, and similarity computation was done on these vectors to find duplicates. They reported a recall rate of ~30\% for k=5 and a search space of bug reports with age $< 100$ days. \cite{b2}.  

In reference \cite{b5}, Neysiani and Babamir contrasted Information Retrieval and Machine Learning based approaches. The study observed that Information Retrieval methods tend to return a list of potential duplicate bug reports based on the similarity between the existing and new bug reports. While the ML classification-based approach usually trains on pairs of identical and non-identical bug reports. The results for retrieval-based strategies are traditionally calculated using the Mean Average Precision and Recall rate. Whereas the test results for ML classification are calculated using accuracy, precision, recall, f1-score, etc. The study also reports that while the accuracy reported for classification-based models is high, they are rarely used in real-world bug tracking systems. This is because it requires two bug reports to predict whether input reports are duplicates of each other. On the other hand, information retrieval-based models are more useful as they report a list of existing originals that are identical to the new bug report.  

A. T. Nguyen et al. \cite{b6} used a hybrid approach combining information retrieval and topic modeling. The study reported an accuracy improvement of 20\%. It used LDA for topic modeling bug reports. Topic modeling for bug reports considers the probability of a word belonging to a particular topic. This works well with bug reports using different terms describing the same issue. By using BM25F and T-Model, Textual and Topic similarity is measured to detect existing duplicate bug reports.  

Haruna et al. Fine-tuned a pre-trained BERT triplet network that uses pairs. Duplicate-Original bug reports made the positives class, and Original-Unrelated bug reports made the negative class. The original bug report was the same (anchor) in both positive and negative cases \cite{b7}.  

J. Lerch and M. Mezini proposed using stack traces available from execution to run a search for existing bug reports with similar stack trace information in them. Thus, saving the user the time to go through the hassle of filling out the whole bug report only to find that it is a duplicate report \cite{b4}.  

Xiaoyin Wang et al. Used both the bug information like summary and description along with the execution information to calculate similarities between two bug reports. Using execution information is a interesting approach. They try to replicate the bug using steps to reproduce and track the functions which get called during the execution. They use this information to build a second set of vectors and aggregate similarity scores from bug report information as well as execution information. Their experiment achieved a ~90\% recall rate for k=5 on an extensively small dataset of only 232 bug reports with only 42 pairs of duplicates \cite{b9}.  

Ashima Kukkar et al. Use CNN for feature extraction, a deep learning model that captures the syntactic and semantic meaning of features used for similarity computations. The study utilizes several openly available bug data sources, for example, Eclipse, Open Office, Firefox, etc. The paper reports 89\%,83\%, and 78\% recall rates for k=5 for Eclipse, Open Office, and Firefox, respectively. The formula for the recall is based on the classification approach. It also accounts for the correct classification of non-duplicate reports, which comprise 80\% to 90\% of the data \cite{b10}.  

Alipour, Hindle and Stroulia, Incorporated more features by inference from bug reports apart from the information already available. They feature engineered contextual features to improve duplicate bug retrieval. They created 6 contexts, each linked to a list of most used words to describe the context. When computing similarity between two bug reports, they use vectors from these new features along with the typically available details from a bug report \cite{b11}.  

B. Kucuk and E. Tuzun characterized Master, Duplicate, and Unique bug reports. Also identified traits of resolved master-duplicate bug reports and unresolved master-duplicate bug reports and proposed actions that may be taken for each category \cite{b12}. The proposed actions seem pragmatic as they suggest re-opening the resolved existing bug report or creating a new scope if the bug tracking system allows it. For the Master bug report that is still open, the new bug report can be prevented from being submitted. They also identified differences in Parent and Child bug reports in the above two categories based on characteristics such as severity, users involved, bug surface time, and life cycle time.  

Rocha and Carvalho, proposed SiameseQAT, which considers reports for individual bugs as well as information from clusters of Parent-Child Reports where each parent can have multiple children. Thus, the Parent Report becomes the centroid of the collection. SiameseQAT combines syntactical and semantical learning on structured, unstructured, and topic extraction-based features as well. They reported results on Eclipse, Netbeans, and Open Office datasets and reported a recall@25 mean of 85\% \cite{b13}.  

H. Mahfoodh and M. Hammad used Word2Vec and Tensorflow to build a neural network-based machine learning model to identify duplicate bug reports. They use a word matching algorithm to determine the similarities between two sentences \cite{b14}.  

In reference \cite{b15} Xiao, Du, Sui and Yue, introduced a novel heterogeneous information network is used to incorporate structured details in addition to unstructured characteristics of a bug report, for example, version, priority, severity, etc. The HIN combines representation learning to develop a deep neural network, hence HINDBR. They report a high 98\% classification accuracy on a small dataset made of pairs of duplicate and non-duplicate bug reports \cite{b15}. 

\section{Our Approach}

\begin{figure}[!h]
\centerline{\includegraphics[width=0.5\textwidth]{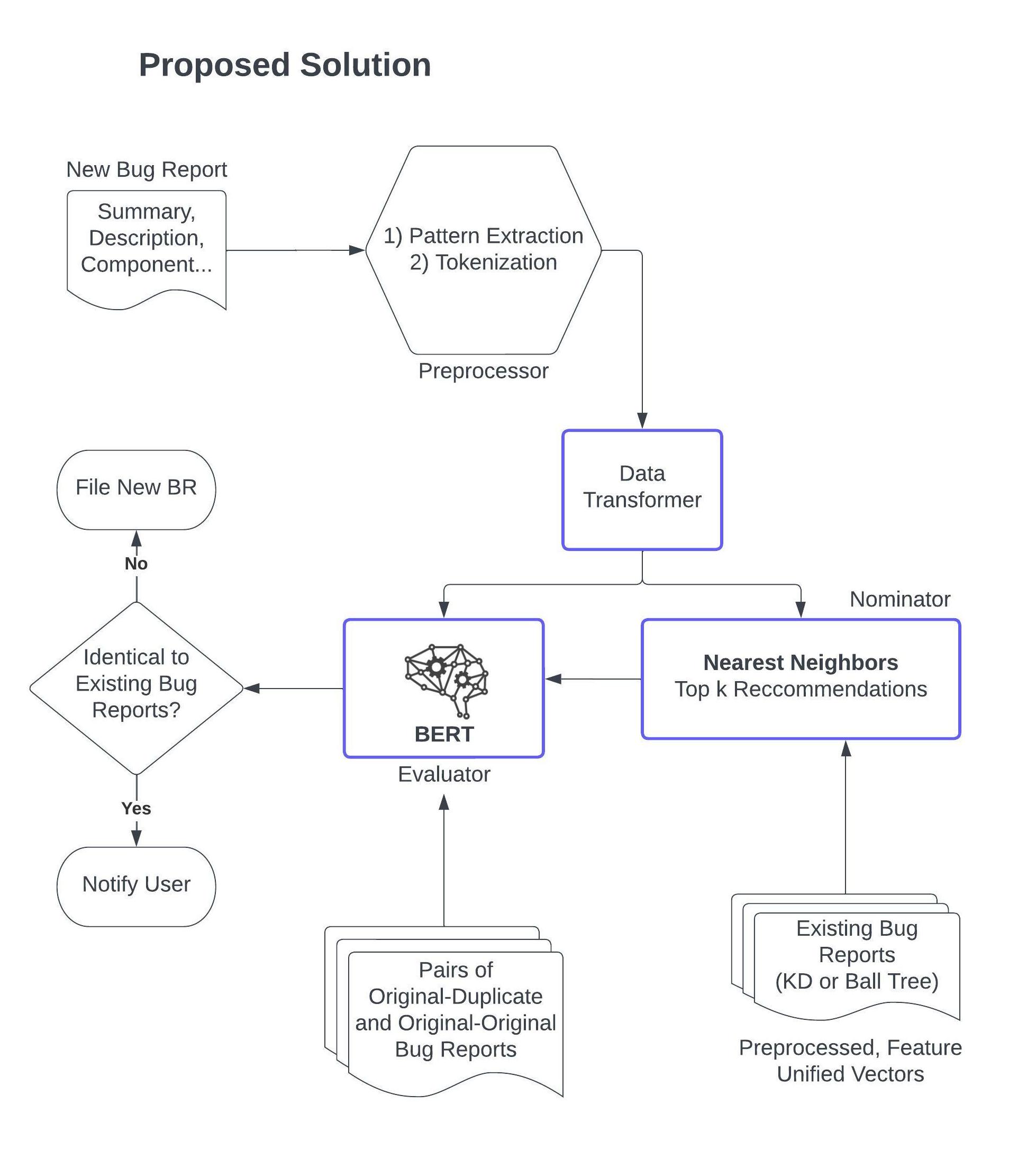}}
\caption{Proposed Solution Flowchart}
\label{fig:arch}
\end{figure}

In this research, we have built a similarity search search mechanism by leveraging NLP approaches that interact with each other to justify the extent to which a Bug Report is similar to an existing one. Our system is made of four components; Preprocessor, Data Transformer, Nominator, and Evaluator. As outlined in Fig \ref{fig:arch}.

\subsubsection{Preprocessor}
As part of data preprocessing techniques, we first applies basic preprocessing techniques to clean the data by tokenizing the unstructured text-based information followed by some data domain relevant feature engineering steps such as in our usecase, the alpha or alphanumeric are considered features. We use the alphanumeric terms as these features can refer to vital information that helps bug identification. For example, an alphanumeric script name is provided as a part of a stack trace in the description of a bug report. However, stack traces can also have process identification numbers or hexadecimal memory addresses, which do not contribute specific information that helps triage a bug report. Furthermore, it will try to find execution failure traces in the description of the problem \cite{b4}\cite{b9}. If found, the preprocessor adds the extracted stack trace to the bug report summary again to increase its weightage in learning \cite{b2}. Identical bug reports can have the same values for specific characteristics like symptoms, impact, and priority \cite{b12}, considering these data is useful for the model \cite{b15}. If present these characteristic data are concatenated to Summary. Not using them as independent features helps reduce mismatches for reports where the values for these attributes are not provided. The processed corpus or document is then fed to the Data Transformer. 

\subsubsection{Data Transformer}
 The Data Transformer component employs two methods to transform the text documents into vectors. The first one is Scikit Learn’s Column Transformer which acts as a wrapper for distinct types of data transformers. It allows the aggregation of heterogeneous data as a single vector. For example, the column transformer used here is made of TFIDF Vectorizer, Dictionary Vectorizer, and a One Hot Encoder. Bug report attributes like Summary and Description are transformed using TFIDF, while Component can be vectorized using OHE. The Dictionary Vectorizer can be used for bug reports with structured data in a dictionary format, for example, Platforms. The column transformer also supports weight assignment. The Summary, Description, Component, and Platform data transformers have weights of 45\%, 25\% \cite{b8}, 25\%, and 5\%, respectively. These transformations are used to build the Nearest Neighbors model. The second method employed by Data Transformer simply concatenates the earlier mentioned characteristics of the bug report and uses Input Example from the Sentence Transformer library. These transformations are used in pairs to train the BERT model.

\subsubsection{Nominator}
The Nominator utilizes unsupervised nearest neighbors learning. It can select from three nearest neighbor algorithms: Brute Force, KD Tree, or Ball Tree. The nominator algorithm attempts to determine the best approach from the training data. The choice depends on factors like the number of samples, the number of features, if the data is sparse or not, the number of neighbors requested for each query, and the number of queries. If the number of samples and searches is large, brute force would be a wrong choice, but if the amount of data and inquiries is small, we can avoid the overhead of constructing the KD or Ball Trees. The input to The Nominator component is the vectors from the Column Transformer. For training, the Nominator would just determine the nearest neighbor algorithm it needs to use and construct the Trees if required. Which does not take more than a few minutes. We also maintain a cache of recently submitted unique bug reports. This helps us detect identical submissions in case a duplicate report is filed within a few minutes. Thus, we ensure that model knowledge is current up to the latest bug report. For Information Retrieval, the Nominator will return a list of nearest neighbors along with the Euclidean distance from them. The similarity is then computed by subtracting distance from the unity. This list of nearest neighbors contains indices that can be used to get the Bug Report ID. We use these IDs to load, preprocess and transform the nominated bug reports using the Input Example transformer mentioned in the data transformer component. These vectors are sent to the Evaluator for further assessment.
\subsubsection{Evaluator}
The evaluator uses BERT, a deep neural network, as a filter. BERT is trained differently compared to unsupervised nearest neighbors learning used in the Nominator. To train BERT we use pairs of known original-duplicate and original-original bug reports, also known as the Siamese network. The original-duplicate couple makes the positive case, while the original-original couple is a negative case. For training BERT, we considered three times the data samples of the positives as compared to the negatives. This helps reflect the nature of data and the fact that fewer duplicate bug reports are filed compared to unique bug reports. Furthermore, this allows BERT to learn a distinct set of patterns from the Nearest Neighbors model. Internally, BERT makes use of Transformer, an attention mechanism that learns contextual relations between words (or sub-words) in a text. Based on this, the BERT model evaluates predictions from unsupervised nearest neighbors. A New bug report is paired with each recommendation from the model to assess whether they are identical or not. Suppose the prediction is identified as a Negative case in BERT, a new bug report paired with a similar bug report being an original-original pair. In that case, it will suppress that recommendation from the Nominator. This suppression applied by BERT eliminates False Positives resulting in a shorter precise list of existing identical bug reports. 

\section{Evaluation \& Results}
\subsection{Dataset}
\begin{enumerate}
    \item \textbf{Private Dataset:} A collection of bug reports from a private enterprise, It has 100+ components and 60K+ bug reports.
    \item \textbf{Firefox Dataset:} A collection of bug reports for the open-source web browser Firefox. It has 50+ components and 100K+ bug reports.
    \item \textbf{Eclipse Dataset:} A collection of bug reports for the open-source Integrated Development Environment. It has 20+ components and 80K+ bug reports.
\end{enumerate}
\subsection{Evaluation Metrics}
 For Information Retrieval systems, recall rate accuracy is typically important performance evaluation metrics \cite{b3}. Recall rate is the percentage of duplicates for which the original (also known as parent or master) bug report is found in a list of size $n$ \cite{b1}\cite{b2} expressed as below: 
\begin{equation}
Recall(n)={\sum_{i=1\ldots \#duplicates}listed(n)\over \# actual\ duplicates}
\label{eq:1}
\end{equation}
 
 The list presents identical bug reports that exist in the bug tracking system. For a search of duplicates, where the size of the list of potentially identical bug reports is minimal, i.e., k=5, we call it a successful search if the Original or Parent bug report is present in the list. The recall rate \cite{b1} thus helps us measure the efficiency of similarity search tools. We would also like to see the parent report at or near the top of the list. The total percentage of bug report from test data for which correct known parent report is found in the list measures the accuracy of the model. 

 We use the Euclidean Distance to measure the similarity between two bug reports. If the distance between two data samples is zero, the samples are identical. Greater distance between two documents lowers the similarity. Euclidean Distance calculation can be expressed as following:

\begin{equation}
Distance(p,q) = \sqrt {\sum _{i=1}^{d}  \left( q_{i}-p_{i}\right)^2 }
\label{eq:2}
\end{equation}
\newline
Where, $p,q$ are two points in Euclidean d-space, $q_{i},p_{i}$ are euclidean vectors, starting from the origin of the space (initial point), and $d$ = d dimensional space.
Similarity can be calculated by subtracting the distance from the unity for the data samples for which space is zero; the resemblance is 1 or 100\% in percentages expressed by (\ref{eq:3}) below:
 
\begin{equation}
Similarity(p,q) = 1-\sqrt {\sum _{i=1}^{d}  \left( q_{i}-p_{i}\right)^2 }
\label{eq:3}
\end{equation}
 
 We consider three significant data sources (private as well as public) containing thousands of bug reports, one private and two public. Table \ref{table:ttdc} is the count of bug reports used in training and testing the model. Experiments indicate that the proposed solution achieves a high average accuracy of 70\% for k=5, i.e., recall@5 across all data. 0This means that the original is found in a list recommending 5 similar bug reports that already exist in the bug tracking system for 70\% of duplicate bug reports. The reported results could be conservative as the recall rate can be higher. This happens because there are duplicate bug reports in the system which are allowed to exist as Unique reports.


\begin{table}
\caption{Training \& Testing data count}
\centering
\begin{tabular}{|c|c|c|}
        \hline
        Dataset & Train Count & Test Count \\
        \hline
        Firefox &90963 &9332 \\
        Eclipse &71131 &6543\\
        Private &61396 &4680\\
        \hline
\end{tabular}
\label{table:ttdc}
\end{table}

\subsection{Results}
The bug reports used are the ones that are known duplicates and each corresponding Parent bug report. For each known Child bug report, we search for a Parent bug report. We also keep track of a couple of things. First, if the parent is present in the list or not. Second, the index of the parent if found in the list, and lastly, the similarity percentage between the Child and the Parent bug report. Following are the results per data set. 

\subsubsection {For Private Dataset}

\begin{figure}[htbp]
\centerline{\includegraphics[width=0.5\textwidth]{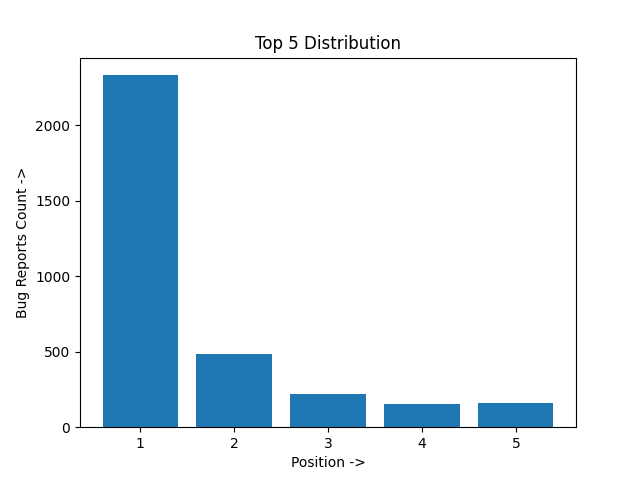}}
\caption{Identical Bug Report rang in Top 5 Recommendations, Private.}
\label{fig:pvt1}
\end{figure}

\begin{figure}[htbp]
\centerline{\includegraphics[width=0.5\textwidth]{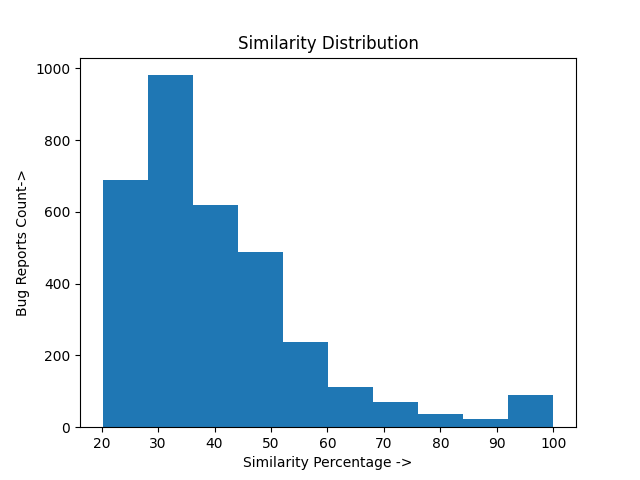}}
\caption{Distribution of Similarity for internal bug reports data.}
\label{fig:pvt2}
\end{figure}

\begin{figure}[htbp]
\centerline{\includegraphics[width=0.5\textwidth]{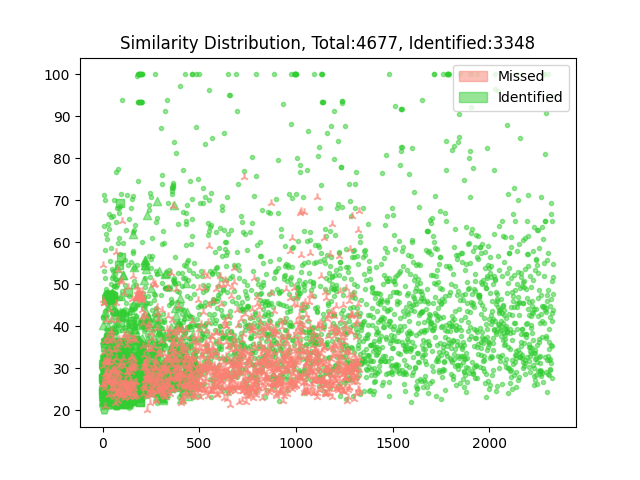}}
\caption{71.58\% Parent BRs in Top 5 for internal data.}
\label{fig:pvt3}
\end{figure}

We observed that the Parent bug report was the top recommendation for 2332 Child reports. The distribution of position and count of bug reports, similarity, and accuracy is represented by the charts Fig \ref{fig:pvt1}, Fig \ref{fig:pvt2}, Fig \ref{fig:pvt3}.

\subsubsection{For Firefox Dataset} 

\begin{figure}[htbp]
\centerline{\includegraphics[width=0.5\textwidth]{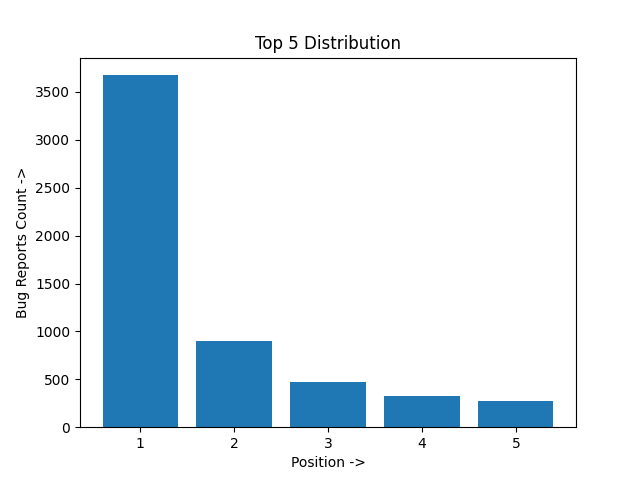}}
\caption{Identical Bug Report rang in Top 5 Recommendations, Firefox.}
\label{fig:ff1}
\end{figure}

\begin{figure}[htbp]
\centerline{\includegraphics[width=0.5\textwidth]{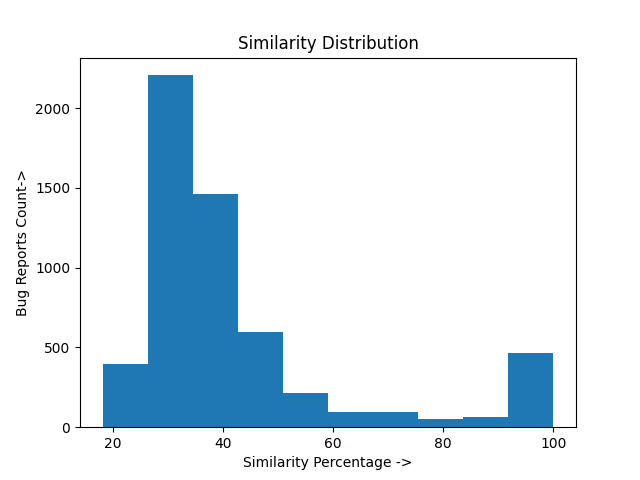}}
\caption{Distribution of Similarity for Firefox bug reports data.}
\label{fig:ff2}
\end{figure}

\begin{figure}[htbp]
\centerline{\includegraphics[width=0.5\textwidth]{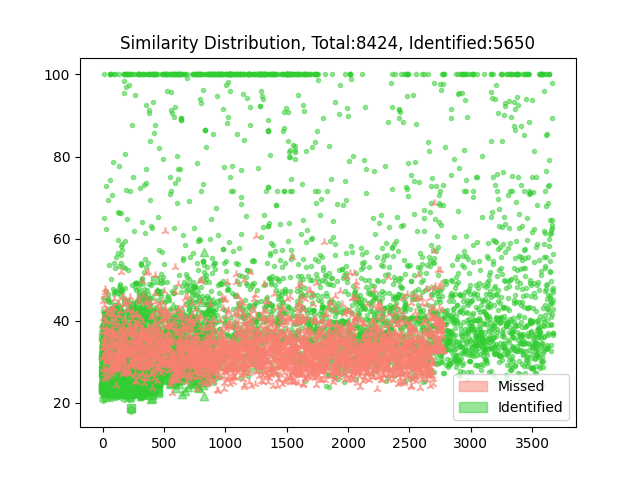}}
\caption{67\% Parent BRs in Top 5 for Firefox data.}
\label{fig:ff3}
\end{figure}

The Parent bug report was the top recommendation for this dataset for 3675 Child reports. The distribution of position and count of bug reports, similarity, and accuracy is represented by the charts Fig \ref{fig:ff1}, Fig \ref{fig:ff2}, Fig \ref{fig:ff3}.

\subsubsection {For Eclipse Dataset}

\begin{figure}[htbp]
\centerline{\includegraphics[width=0.5\textwidth]{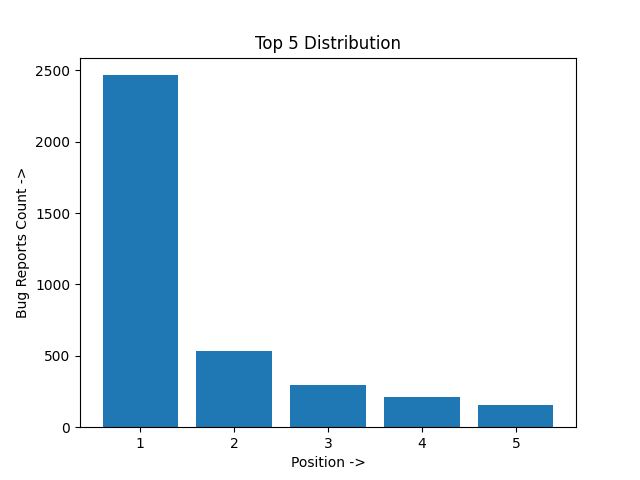}}
\caption{Identical Bug Report rang in Top 5 Recommendations, Eclipse.}
\label{fig:ec1}
\end{figure}

\begin{figure}[htbp]
\centerline{\includegraphics[width=0.5\textwidth]{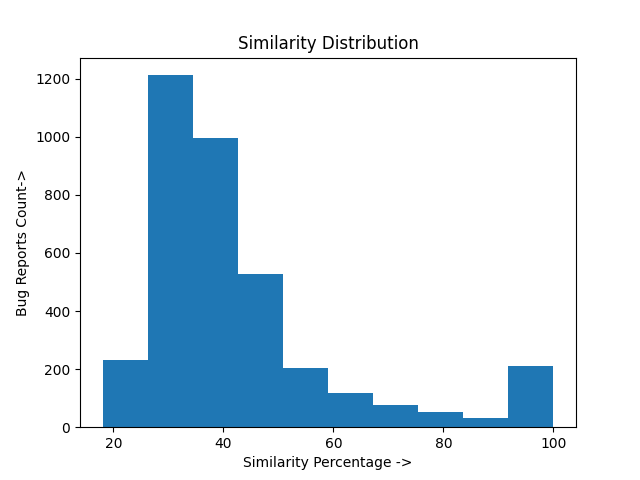}}
\caption{Distribution of Similarity for Eclipse bug reports data.}
\label{fig:ec2}
\end{figure}

\begin{figure}[htbp]
\centerline{\includegraphics[width=0.5\textwidth]{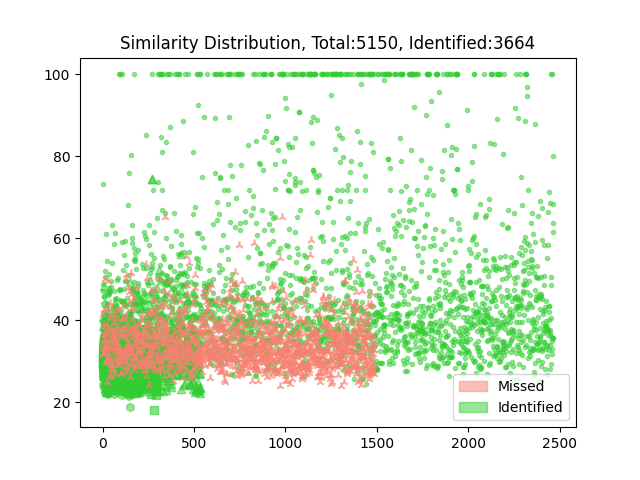}}
\caption{71.14\% Parent BRs in Top 5 for Eclipse data.}
\label{fig:ec3}
\end{figure}

The Parent bug report was the top recommendation for 2466 Child reports. The distribution of position and count of bug reports, similarity, and accuracy is represented by the charts Fig \ref{fig:ec1}, Fig \ref{fig:ec2}, Fig \ref{fig:ec3}.

\begin{figure}[htbp]
\centerline{\includegraphics[width=0.5\textwidth]{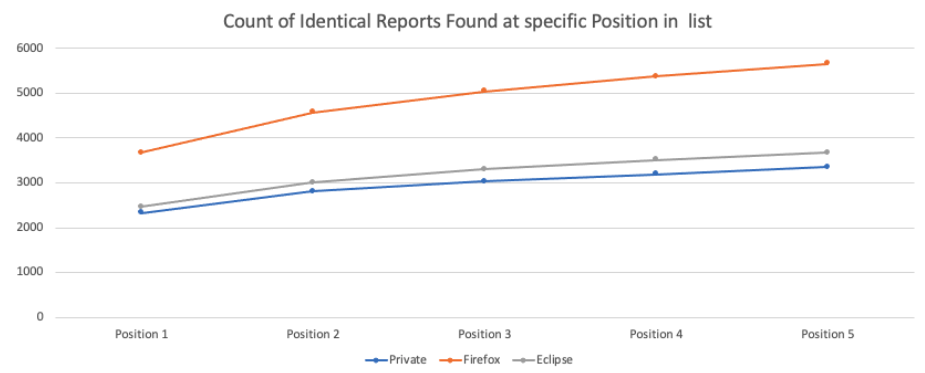}}
\caption{Cumulative Count vs. Position of Identified Parent Report}
\label{fig:ovl}
\end{figure}

The overall performance of the proposed solution is depicted in the Table \ref{table:ovlpf}. The left column represents the position where the parent bug report was found. The numbers in each row show the count of bug reports per data set found in the corresponding place. -1 is the count of bug reports which had no parent in the predicted list from model. Fig \ref{fig:ovl} represents Cumulative Count vs. Position of Identified Parent Report. \newline

\begin{table}
\caption{Position and Count of Identified Parent Bug Reports}
\centering
\begin{tabular}{|c|c|c|c|}
        \hline
        Position & Private & Firefox & Eclipse \\
        \hline
        1 &2332 &3675 &2466\\
        2 &482 &901 &535\\
        3 &219 &469 &295\\
        4 &157 &328 &209\\
        5 &158 &277 &159\\
        -1 &1329 &2774 &1486\\
        \hline
        Total &4677 &8424 &5150\\
        \hline
\end{tabular}
\label{table:ovlpf}
\end{table}

\section{Limitations}
The results rely on the correctness of the data evaluated. We know that a bug tracking system may not have all the duplicate bug reports identified and marked, which restricts the performance calculation of the model. Also, some child reports do not directly link to a parent report because they map to a different child of the same parent report \cite{b13}. However, these discrepancies only affect the score of the model, not the model itself. The model is able to weed out the duplicate reports that existed in the system as originals. In addition, the model links child reports to the correct parents as the Nearest Neighbor's model does not contain duplicate bug reports in its knowledge base.

\section{Conclusion}
Using deep neural network with BERT as the evaluator of Nearest Neighbors Unsupervised learning, and an enhanced weighted heterogeneous data transformer is a novel idea that helps us get high recall rate accuracy. Independently, the nearest neighbor learning, which uses Term Frequency-based vector, fails to achieve equivalent results \cite{b2}\cite{b6}. It cannot learn the relationship among the words in the Summary and Description of the bug report. This is because TF-IDF does not consider the semantic meaning or context of the words\cite{b17}, whereas BERT does. On the other hand, BERT does not consider the all the tokens in the document like TFIDF does, but BERT can derive the contexts of the terms and understand their importance \cite{b16}\cite{b7}. Since BERT uses deep neural networks as part of its architecture \cite{b16}, it can be computationally expensive compared to Nearest Neighbors. BERT also limits the number of tokens it considers to 512, which Nearest Neighbors does not. Put together, we overcome the deficiencies of individual models. We get the best from each, Weighted frequencies from TFIDF, fast search of Nearest Neighbors, and Contextual understanding from BERT.


\begin{thebibliography}{00}
\bibitem{b1} J. Natt och Dag, V. Gervasi, S. Brinkkemper and B. Regnell, "Speeding up requirements management in a product software company: linking customer wishes to product requirements through linguistic engineering," Proceedings. 12th IEEE International Requirements Engineering Conference, 2004., 2004, pp. 283-294, doi: 10.1109/ICRE.2004.1335685.
\bibitem{b2} P. Runeson, M. Alexandersson and O. Nyholm, "Detection of Duplicate Defect Reports Using Natural Language Processing," 29th International Conference on Software Engineering (ICSE'07), 2007, pp. 499-510, doi: 10.1109/ICSE.2007.32.
\bibitem{b3} Manning, C. D. and Schütze, H., Foundations of Statistical Natural Language Processing. Cambridge, USA: MIT Press, 1999.
\bibitem{b4} J. Lerch and M. Mezini, "Finding Duplicates of Your Yet Unwritten Bug Report," 2013 17th European Conference on Software Maintenance and Reengineering, 2013, pp. 69-78, doi: 10.1109/CSMR.2013.17.
\bibitem{b5} B. S. Neysiani and S. Morteza Babamir, "Automatic Duplicate Bug Report Detection using Information Retrieval-based versus Machine Learning-based Approaches," 2020 6th International Conference on Web Research (ICWR), 2020, pp. 288-293, doi: 10.1109/ICWR49608.2020.9122288.
\bibitem{b6} A. T. Nguyen, T. T. Nguyen, T. N. Nguyen, D. Lo and C. Sun, "Duplicate bug report detection with a combination of information retrieval and topic modeling," 2012 Proceedings of the 27th IEEE/ACM International Conference on Automated Software Engineering, 2012, pp. 70-79, doi: 10.1145/2351676.2351687.
\bibitem{b7} H. Isotani, H. Washizaki, Y. Fukazawa, T. Nomoto, S. Ouji and S. Saito, "Duplicate Bug Report Detection by Using Sentence Embedding and Fine-tuning," 2021 IEEE International Conference on Software Maintenance and Evolution (ICSME), 2021, pp. 535-544, doi: 10.1109/ICSME52107.2021.00054. 
\bibitem{b8} T. Hirsch and B. Hofer, "Identifying non-natural language artifacts in bug reports," 2021 36th IEEE/ACM International Conference on Automated Software Engineering Workshops (ASEW), 2021, pp. 191-197, doi: 10.1109/ASEW52652.2021.00046.
\bibitem{b9} X. Wang, L. Zhang, T. Xie, J. Anvik and J. Sun, "An approach to detecting duplicate bug reports using natural language and execution information", Proc. 13th Int. Conf. Softw. Eng., pp. 461-470, 2008. 
\bibitem{b10} A. Kukkar, R. Mohana, Y. Kumar, A. Nayyar, M. Bilal and K. -S. Kwak, "Duplicate Bug Report Detection and Classification System Based on Deep Learning Technique," in IEEE Access, vol. 8, pp. 200749-200763, 2020, doi: 10.1109/ACCESS.2020.3033045.
\bibitem{b11} A. Alipour, A. Hindle and E. Stroulia, "A contextual approach towards more accurate duplicate bug report detection," 2013 10th Working Conference on Mining Software Repositories (MSR), 2013, pp. 183-192, doi: 10.1109/MSR.2013.6624026. 
\bibitem{b12} B. Kucuk and E. Tuzun, "Characterizing Duplicate Bugs: An Empirical Analysis," 2021 IEEE International Conference on Software Analysis, Evolution and Reengineering (SANER), 2021, pp. 661-668, doi: 10.1109/SANER50967.2021.00084. 
\bibitem{b13} T. M. Rocha and A. L. D. C. Carvalho, "SiameseQAT: A Semantic Context-Based Duplicate Bug Report Detection Using Replicated Cluster Information," in IEEE Access, vol. 9, pp. 44610-44630, 2021, doi: 10.1109/ACCESS.2021.3066283.
\bibitem{b14} H. Mahfoodh and M. Hammad, "Word2Vec Duplicate Bug Records Identification Prediction Using Tensorflow," 2020 International Conference on Innovation and Intelligence for Informatics, Computing and Technologies (3ICT), 2020, pp. 1-6, doi: 10.1109/3ICT51146.2020.9311954. 
\bibitem{b15} G. Xiao, X. Du, Y. Sui and T. Yue, "HINDBR: Heterogeneous Information Network Based Duplicate Bug Report Prediction," 2020 IEEE 31st International Symposium on Software Reliability Engineering (ISSRE), 2020, pp. 195-206, doi: 10.1109/ISSRE5003.2020.00027.
\bibitem{b16} Devlin, Jacob, Ming-Wei Chang, Kenton Lee, and Kristina Toutanova. “Bert: Pre-Training of Deep Bidirectional Transformers for Language Understanding.” arXiv.org, May 24, 2019.
\bibitem{b17} M. Li and B. -B. Yin, "ARB-BERT: An Automatic Aging-Related Bug Report Classification Method based on BERT," 2021 8th International Conference on Dependable Systems and Their Applications (DSA), 2021, pp. 474-483, doi: 10.1109/DSA52907.2021.00071.
\bibitem{b18} L. Wang, L. Zhang and J. Jiang, "Detecting Duplicate Questions in Stack Overflow via Deep Learning Approaches," 2019 26th Asia-Pacific Software Engineering Conference (APSEC), 2019, pp. 506-513, doi: 10.1109/APSEC48747.2019.00074. 
\bibitem{b19} Z. Chen, X. Ju, Y. Shen and X. Chen, "Improving Blocking Bug Pair Prediction via Hybrid Deep Learning," 2021 IEEE 21st International Conference on Software Quality, Reliability and Security Companion (QRS-C), 2021, pp. 727-732, doi: 10.1109/QRS-C55045.2021.00110.
\bibitem{b20} S. K. Chauhan, A. Goel, P. Goel, A. Chauhan and M. K. Gurve, "Research on product review analysis and spam review detection," 2017 4th International Conference on Signal Processing and Integrated Networks (SPIN), 2017, pp. 390-393, doi: 10.1109/SPIN.2017.8049980.


\end{thebibliography}
\end{document}